\documentstyle[aps,psfig,epsfig]{revtex}

\newcommand{\ep}{\epsilon}

\newcommand{\pa}{\partial}

\newcommand{\td}{\tilde}
\newcommand{\sla}[1]{\slash\!\!\! #1}

\begin{document}

\draft

\title{\hspace{14cm}{\rm {\normalsize{USTC-ICTS-0301}}}\\
\vspace{1cm}
Resummation Study on Decay $\rho\rightarrow\pi\pi$ in
$U(2)_L\times U(2)_R$ Chiral Theory of Mesons}
\author{Yi-Bin Huang\footnote{E-mail address:
huangyb@mail.ustc.edu.cn}}
\address{Interdisciplinary Center for Theoretical Study,
University of Science and Technology of China\\
Hefei, Anhui 230026, P. R. China}
\author{Mu-Lin Yan\footnote{E-mail address: mlyan@staff.ustc.edu.cn}}
\address{CCST(World Lab), P. O. Box 8730, Beijing, 100080, P. R. China \\
  and\\
 Interdisciplinary Center for Theoretical Study,
University of Science and Technology of China\\
Hefei, Anhui 230026, P. R. China\footnote{mail address}}
\author{Xiao-Jun Wang\footnote{E-mail address: wangxj@itp.ac.cn}}
\address{Institute of Theoretical Physics, Beijing, 100080, P. R. China}
\date{\today}
\maketitle

\begin{abstract}
We improve $O(p^4)$ calculation in $U(2)_L\times U(2)_R$ chiral
theory of mesons by resummation calculation for vector mesons
physics and restudy decay $\rho\rightarrow\pi\pi$. A complete and
compact expression for $f_{\rho\pi\pi}(p^2)$ (up to $O(p^\infty)$)
is obtained, from which an important non-perturbative conclusion
is given based on convergence and unitarity consideration.
\end{abstract}


The Nambu-Jona-Lasinio (NJL) model \cite{Nambu} and its extensions
are widely used to understand hadron physics (for recent reviews
see e.g. \cite{Bijnens}). There are various methods to parametrize
the NJL version models, among which is the $U(N_f)_L\times
U(N_f)_R$ chiral theory of mesons \cite{Li95} with $N_f=2\;{\rm
or}\;3$ (proposed by Li and called hereafter Li's model). This
model can be regarded as a realization of chiral symmetry, current
algebra and vector meson dominance (VMD). It provides a unified
description of pseudoscalar, vector, and axial-vector mesons,
which are introduced as bound states of quark fields. The basic
inputs of it are the cutoff $\Lambda$ (or $g$ in \cite{Li95}) and
constituent quark mass $m$ (related to quark condensate). This
theory has been studied extensively
\cite{Li97,LiNucl97,GaoYan,WangChi,LiCoe,GaoYanRho}, in particular
it has been used recently \cite{Li02} to analyze the data of g-2
of muon reported by CMD-2 group \cite{CMD}. Hence Li's model is a
good phenomenological model.

So far Li's model is only investigated up to $O(p^4)$ \cite{Li95}
in perturbative manner. This treatment is reasonable for chiral
perturbation theory (ChPT) \cite{GL85a}, because the typical
energy there is much less than the energy scale of chiral symmetry
spontaneously breaking (CSSB) $\Lambda_{\rm CSSB}\sim 2\pi F_\pi
\sim 1$GeV. This is not the case here because the typical energy
$p$ of vector mesons is comparable with $\Lambda_{\rm CSSB}$. For
example, at $\rho$ energy scale, $p\sim m_\rho$ and the
perturbation parameter $p/\Lambda_{\rm CSSB}\sim0.7$. Therefore,
studies in ref. \cite{Li95} should be improved by going beyond
$O(p^4)$. But we know that, in usual perturbative treatment, it is
difficult to deduce effective meson action at higher order. This
is a paradox. As a solution to it, we develop a method to obtain
the sum of all terms of the $p$-expansion \cite{WY00}. We call it
resummation study. In this paper, we want to use it to restudy
typical decay $\rho\rightarrow\pi\pi$ in Li's model. Other
processes can be restudied in a similar way.

Before this study, let us outline procedures of perturbative
calculation and our resummation calculation. Given a quark model
which describes interactions of constituent quarks $q$ and mesons
$M$ and satisfies chiral symmetry requirement, we can write the
Lagrangian as ${\cal L}={\cal L}(q,\bar{q},M)=\bar{q}{\cal D}q$,
where ${\cal D}={\cal D}(M)$ is characteristic of the model. The
first step of perturbative calculation is using the method of path
integral to integrate out the quark fields:
\begin{eqnarray}\label{1.1}
e^{i\int d^4x{\cal L}_{\rm eff}(M)}=\int[d q][d\bar{q}] e^{i\int
d^4x\bar{q}{\cal D}q}.
\end{eqnarray}
After functional integration, the effective action $S_{\rm
eff}(M)$ of mesons is formally obtained as $S_{\rm eff}(M)={\rm
ln}\;{\rm det}{\cal D}(M)$. To regularize this determinant,
Schwinger's proper time method \cite{Sch54} or heat kernel method
\cite{Ball89} should be employed. This directly results in an
expansion in $p$, and usually up to $O(p^4)$. Attempting to
calculate terms of higher orders will encounter great difficulty.

On the other hand, to perform resummation study, we calculate
effective meson action via loop effects of constituent quarks.
This is equivalent to integrating out quarks in path integral, but
via it we can obtain effects from all orders of $p$-expansion.
Specifically, we divide ${\cal L}$ into free-field part ${\cal
L}_0$ and interaction part ${\cal L}_{\rm I}$, and turn to
interaction picture. The effective action can be obtained as
\begin{eqnarray}\label{3.3}
e^{iS_{\rm eff}}&=&<0|T_qe^{i\int d^4x{\cal L}_{\rm I}}|0>
       \nonumber \\
 &=&\sum_{n=1}^\infty i\int d^4p_1\frac{d^4p_2}{(2\pi)^4}
  \cdots\frac{d^4p_n}{(2\pi)^4}\delta^4(p_1+p_2+\cdots+p_n)\Pi_n(p_1,\cdots,p_n),
\end{eqnarray}
where $T_q$ is time-order product of constituent quark fields,
$\Pi_n(p_1,\cdots,p_n)$ is one-loop effects of constituent quarks
with $n$ external fields, $p_1,p_2,\cdots,p_n$ are their
four-momentum. Getting rid of all disconnected diagrams, we have
\begin{eqnarray}\label{3.5}
S_{\rm eff}=\sum_{n=1}^\infty S_n, \hspace{0.6in} S_n=\int
d^4p_1\frac{d^4p_2}{(2\pi)^4}\cdots\frac{d^4p_{n}}
  {(2\pi)^4}\delta^4(p_1+p_2+\cdots+p_n)\Pi^c_n(p_1,\cdots,p_{n}),
\end{eqnarray}
where $c$ denotes connected part. Obviously, in eq.~(\ref{3.5}),
the effective action $S_{\rm eff}$ is expanded in number of
external vertex and expressed as integral over external momentum.
Hereafter we shall call this method proper vertex expansion, and
call $S_n$ $n$-point effective action. In terms of proper vertex
expansion, the effective actions include informations from all
orders of chiral expansion. That is, we can do resummation of
momentum expansion by this method. That is what we need.

It is instructive to compare these two kinds of calculations. The
perturbative calculation can deal with all kinds of meson
interactions at low orders, while resummation calculation deals
with specific process but can include informations of all orders.
In the sense of $p$-expansion, resummation calculation is
non-perturbative. As we shall see later, this character will give
important result.

Having showed outlines of resummation calculation, we turn to Li's
model. Because we only focus on $\rho$ physics, the $N_f=2$ part
of Li's model, i.e., $U(2)_L\times U(2)_R$ chiral theory of mesons
\cite{Li95} suffices to study decay $\rho\rightarrow\pi\pi$. This
Model is constructed through $U(2)_L\times U(2)_R$ chiral symmetry
and minimum coupling principle, and the ingredients of it are
quarks ($u$ and $d$), pseudoscalar mesons ($\pi$ and $\eta$ ($u$
and $d$ component)), vector mesons ($\rho$ and $\omega$),
axial-vector mesons ($a_1$ and $f_1(1285)$), lepton, photon and
$W$ bosons. For our purpose,  we just write the relevant
Lagrangian:
\begin{eqnarray}\label{2.1}
{\cal L}=\bar{q}(i\sla{\pa}+\sla{V}+\sla\!{A}\gamma_5-m
u(x))q+\frac{1}{4}m_0^2(<V_\mu V^\mu>+<A_\mu A^\mu>),
\end{eqnarray}
where $A_\mu=\tau_i a_\mu^i+f_\mu$, $V_\mu=\tau_i
\rho_\mu^i+\omega_\mu$, $ u=e^{i\Phi\gamma_5}$, $\Phi=\tau_i
\pi^i+\eta$, and $<\cdots>$ denotes trace in flavor space. The
quark part of this Lagrangian can be divided into two parts. The
free-field part is ${\cal L}^q_0=\bar{q}(i\sla{\pa}-m)q$, and the
interaction part is
\begin{eqnarray}\label{2.2}
{\cal L}_{\rm
I}^q=\bar{q}\big(\sla{V}+\sla\!{A}\gamma_5-im\Phi\gamma_5+\frac{1}{2}m\Phi^2\big)q,
\end{eqnarray}
where terms involving more $\Phi$s has been omitted, because they
have nothing to do with decay $\rho\rightarrow\pi\pi$. All
subsequent calculations are performed at chiral limit.

\begin{figure}[hptb]
   \centerline{\psfig{figure=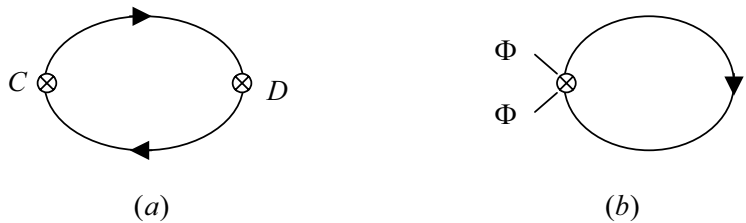,width=4in}}
 \centering
\begin{minipage}{5in}
   \caption{Calculation of $<CD>$ terms ($C,D=V,A,\Phi$) from quark loops.
   a) Two-point diagram of quark loop.
   b) One-point diagram of quark loop for $<\Phi\Phi>$ term.}
\end{minipage}
\end{figure}
Before studying decay $\rho\rightarrow\pi\pi$, we have to do
something to obtain physical fields. We should calculate effective
actions for $<VV>$, $<AA>$, $<\Phi\Phi>$, $<VA>$, $<V\Phi>$ and
$<A\Phi>$ through integration of two-point quark loops like Fig.
1a (where $C,D=V,A,\Phi$) and one-point quark loop Fig. 1b. These
actions are calculated as
\begin{eqnarray}\label{2.3}
S_{\rm eff}^q&=&\frac{iN_c}{2}\int\frac{d^4p
d^4k}{(2\pi2\pi)^4}\nonumber \\
&&\Big(<V_\mu(p)V_\nu(-p)>{\rm Tr}[\gamma^\mu S_F(k-p)\gamma^\nu
S_F(k)]+<A_\mu(p)A_\nu(-p)>{\rm Tr}[\gamma^\mu\gamma_5
S_F(k-p)\gamma^\nu\gamma_5 S_F(k)]\nonumber \\
&&-<\Phi(p)\Phi(-p)>\left(m^2{\rm Tr}[\gamma_5 S_F(k-p)\gamma_5
S_F(k)]+m{\rm Tr}[S_F(k)]\right)\nonumber \\
&&-2im<V_\mu(p)\Phi(-p)>{\rm Tr}[\gamma^\mu S_F(k-p)\gamma_5
S_F(k)]-2im<A_\mu(p)\Phi(-p)>{\rm Tr}[\gamma^\mu\gamma_5
S_F(k-p)\gamma_5 S_F(k)]\nonumber \\
&&+2<V_\mu(p)A_\nu(-p)>{\rm Tr}[\gamma^\mu
S_F(k-p)\gamma^\nu\gamma_5 S_F(k)] \Big),
\end{eqnarray}
where, for $<\Phi\Phi>$ term, we have added contribution from
one-point diagram (Fig. 1b) of quark loop. This is so because
$\Phi$ is realized nonlinearly. After integrating quark loops, we
get expressions including all terms up to $O(p^\infty)$. But we
are not interested in high order terms now. Discarding them and
adding the non-quark part of eq. (\ref{2.1}), we obtain
\begin{eqnarray}\label{2.4}
S_{\rm eff}=\frac{1}{4}\int\frac{d^4p
}{(2\pi)^4}&\Big(&<V_\mu(p)V_\nu(-p)>(g^2(p^\mu
p^\nu-g^{\mu\nu}p^2)+m_0^2)\nonumber \\
&&+<A_\mu(p)A_\nu(-p)>\Big[-g^2\Big(1-\frac{N_c}{6\pi^2g^2}\Big)g^{\mu\nu}p^2+g^2p^\mu
p^\nu+(6g^2m^2+m_0^2)g^{\mu\nu}\Big]\nonumber \\
&&+<\Phi(p)\Phi(-p)>\frac{3}{2}g^2m^2p^2+<A_\mu(p)\Phi(-p)>6ig^2m^2p^\mu\Big),
\end{eqnarray}
where the logarithmic divergence is absorbed by
\begin{eqnarray}\label{2.5}
\frac{3}{8}g^2=\frac{N_c}{(4\pi)^{d/2}}\left(\frac{\mu^2}{m^2}\right)^{\ep/2}
  \Gamma\left(2-\frac{d}{2}\right),\hspace{0.5in}(d=4-\ep)
\end{eqnarray}
and the quadratic divergence in $<\Phi\Phi>$ term has been reduced
to logarithmic divergence by the identity
$\Gamma(1-d/2)+\Gamma(2-d/2)=-1+O(\ep)$. The first and second
terms of eq. (\ref{2.4}) indicate needs for rescalings of $V_\mu$
and $A_\mu$. (We see that axial-vector $A_\mu$ does not satisfy
gauge invariance. Because $\pa_\mu A^\mu=0$ when $A_\mu$ is
on-shell, we shall ignore the $p^\mu p^\nu$ term when redefining
$A_\mu$.) The non-vanishing $<A\Phi>$ term indicates that there is
a mixing between $A_\mu$ and $\pa_\mu\Phi$, which should be erased
by shift of $A_\mu$. Therefore, the redefinitions of $V_\mu$ and
$A_\mu$ are
\begin{eqnarray}\label{2.6}
V_\mu\rightarrow \frac{V_\mu}{g},\hspace{1in}A_\mu\rightarrow
\frac{A_\mu}{g\sqrt{1-N_c/6\pi^2g^2}}-\frac{c}{g}\pa_\mu\Phi,
\end{eqnarray}
and we can obtain in passing the mass formula for them:
\begin{eqnarray}\label{2.7}
m_V^2=\frac{m_0^2}{g^2},\hspace{1in}(1-\frac{N_c}{6\pi^2g^2})m_A^2=6m^2+m_V^2.
\end{eqnarray}
After redefining $A_\mu$ and cancelling the mixing, and then
making the kinetic term of $\Phi$ in standard form, we obtain
$$
c=\frac{3gm^2}{6m^2+m_V^2},\hspace{1in}
\frac{1}{4}c^2m_V^2+\frac{3}{8}\left(1-\frac{2c}{g}\right)^2g^2m^2=\frac{F_\Phi^2}{16},
$$
or
\begin{eqnarray}\label{2.8}
c=\frac{F_\Phi^2}{2gm_V^2},\hspace{1in}6\left(1-\frac{2c}{g}\right)g^2m^2=F_\Phi^2,
\end{eqnarray}
and rescaling for $\Phi\rightarrow 2\Phi/F_\Phi$. Eq.s
(\ref{2.6})-(\ref{2.8}) are exactly those in \cite{Li95}.

\begin{figure}[hptb]
   \centerline{\psfig{figure=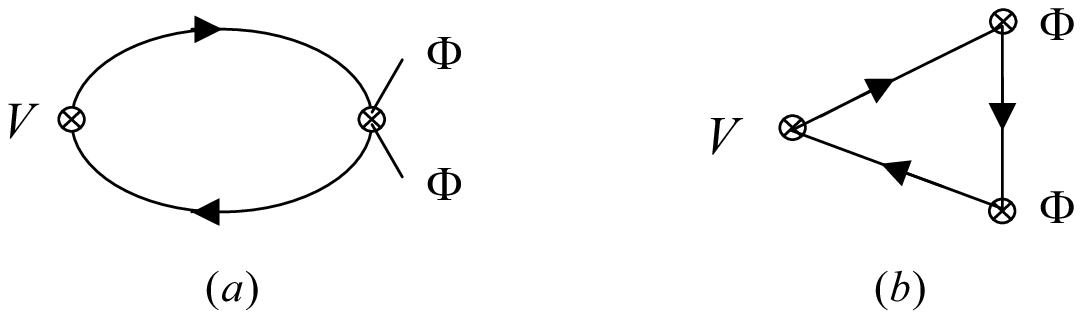,width=4in}}
 \centering
\begin{minipage}{5in}
   \caption{Quark loops for vertex $\rho\rightarrow\pi\pi$. a) Two-point diagram. b) Three-point diagram.}
\end{minipage}
\end{figure}
Now we can study decay $\rho\rightarrow\pi\pi$. For vertex
$\rho-\pi\pi$, we need to calculate two Feynman diagrams of quark
loop (see Fig. 2). The relevant Lagrangians (after redefinitions)
for the first and second ones are
\begin{eqnarray}\label{2.9}
{\cal L}_{\rm
I}^{(2)}=\bar{q}(\frac{1}{g}\sla{V}+\frac{2m}{F_\Phi^2}\Phi^2)q,\hspace{1in}
{\cal L}_{\rm
I}^{(3)}=\bar{q}(\frac{1}{g}\sla{V}-\frac{2c}{gF_\Phi}\sla{\pa}\Phi\gamma_5-\frac{2im}{F_\Phi}\Phi\gamma_5)q
\end{eqnarray}
respectively. For the first one, ${\cal L}_{\rm I}^{(2)}$ gives
integration of the quark loop as $ \int d^4 k{\rm Tr}[\gamma^\mu
S_F(k-p)S_F(k)]\propto p^\mu$ ($p$ is momentum of vector field
$V$). Because $\pa_\mu V^\mu=0$ when $V_\mu$ is on-shell,
contribution from the first diagram vanishes: $S_2=0$. For the
second diagram, the Lagrangian ${\cal L}_{\rm I}^{(3)}$ determines
that the effective action for this three-point diagram is
calculated as
\begin{eqnarray}\label{2.10}
S_3&=&\frac{4iN_c}{gF_\Phi^2}\int\frac{d^4p d^4q_1
d^4q_2}{(2\pi2\pi)^4}\delta^4(p+q_1+q_2)
<V_\mu(p)\Phi(q_1)\Phi(q_2)>\nonumber \\
&&\int\frac{d^4
k}{(2\pi)^4}\Big(\frac{c^2}{g^2}q_{1\nu}q_{2\rho}{\rm
Tr}[\gamma^\mu S_F(k+q_1)\gamma^\nu\gamma_5
S_F(k)\gamma^\rho\gamma_5 S_F(k-q_2)]+m^2{\rm Tr}[\gamma^\mu
S_F(k+q_1)\gamma_5 S_F(k)\gamma_5 S_F(k-q_2)]\nonumber \\
&&\hspace{0.2in}-\frac{cm}{g}q_{1\nu}{\rm Tr}[\gamma^\mu
S_F(k+q_1)\gamma^\nu\gamma_5 S_F(k)\gamma_5
S_F(k-q_2)]-\frac{cm}{g}q_{2\nu}{\rm Tr}[\gamma^\mu
S_F(k+q_1)\gamma_5 S_F(k)\gamma^\nu\gamma_5 S_F(k-q_2)]\Big).
\end{eqnarray}
After integrations of quark loops and considerations of $\pa_\mu
V^\mu=0$ and chiral limit $q_1^2=q_2^2=m_\pi^2=0$, the effective
action for vertex $\rho-\pi\pi$ becomes
\begin{eqnarray}\label{2.11}
S_{\rho\pi\pi}&=&\int\frac{d^4p
d^4q}{(2\pi2\pi)^4}\ep_{ijk}\rho^i_\mu(p)\pi^j(-p-q)\pi^k(q)(-iq^\mu)f_{\rho\pi\pi}(p^2),
\end{eqnarray}
where
\begin{eqnarray}\label{2.12}
f_{\rho\pi\pi}(p^2)&=&\frac{2}{3\pi^2 g
f_\pi^2}\Big[m^2\Big(18(1-2c/g)\pi^2g^2-(2-3c/g)N_c\Big)-p^2(c/g)^2(6\pi^2g^2-N_c)\Big]\nonumber \\
&&+\frac{2N_c}{\pi^2
g f_\pi^2}\int_0^1xdx\int_0^1dy\bigg\{\frac{m^2}{m^2-p^2x(1-x)(1-y)}\Big[m^2(1-2c/g+xy)\nonumber \\
&&\hspace{0.8in}+p^2\Big(x(1-x)(1-y)(1+xy)
-c/g(1+2x-2x^2-3xy+2x^2y)+c^2/g^2(1-xy)\Big)\Big]\nonumber \\
&&\hspace{0.5in}-{\rm
ln}\Big(1-\frac{p^2}{m^2}x(1-x)(1-y)\Big)\Big(m^2(1-4c/g+3xy)-c^2p^2/g^2(1-xy)\Big)
\bigg\}\nonumber \\
&=&\frac{12(N_c+3g^2\pi^2)-(24c/g)(2N_c
+3g^2\pi^2)+40N_c(c/g)^2}{3\pi^2gf_\pi^2}m^2-\frac{c^2(10N_c+36g^2\pi^2)}{9\pi^2g^3f_\pi^2}p^2\nonumber \\
&&-\frac{4N_c\big(3 - 12c/g + 10(c/g)^2\big)m^2 -4N_c
(c/g)^2p^2}{3\pi^2gf_\pi^2}\sqrt{\frac{4m^2 - p^2}{p^2}}{\rm
Arctg}\sqrt{\frac{p^2}{4m^2 - p^2}}.
\end{eqnarray}

The factor $f_{\rho\pi\pi}(p^2)$ is complicated, expansion of it
in $p$ up to $O(p^6)$ (corresponding to expansion of
$S_{\rho\pi\pi}$ up to $O(p^8)$) is
\begin{eqnarray}\label{2.13}
f_{\rho\pi\pi}(p^2)=\frac{2}{g}\Big[1+\frac{N_c(1-2c/g)^2-12c^2\pi^2}{6\pi^2f_\pi^2}p^2+
\frac{(1-4c/g)N_c}{60\pi^2m^2f_\pi^2}p^4
+\frac{\big(1-4c/g+(c/g)^2\big)N_c}{420\pi^2m^4f_\pi^2}p^6+O(p^8)\Big],
\end{eqnarray}
where condition (\ref{2.8}) has been used. As we can see, the sum
of the first two terms in this series is just $f_{\rho\pi\pi}$ in
ref. \cite{Li95} when $N_c=3$ and $p^2=m_\rho^2$ is set.

Moreover, expression (\ref{2.12}) shows that series (\ref{2.13})
is in fact an expansion in dimensionless quantity $\td{p}^2\equiv
p^2/(4m^2)$: $
f_{\rho\pi\pi}(p^2)=2/g+c_1\td{p}^2+c_2\td{p}^4+c_3\td{p}^6+\cdots$.
Convergence of this series means $\td{p}^2<1$ or $p^2<4m^2$, which
at $\rho$ energy scale means the constituent quark mass
$m>m_\rho/2\simeq 385$MeV. The same conclusion can also be
obtained by unitarity consideration. Unitarity of the theory
demands that, at leading order of $1/N_c$ expansion, there should
be no imaginary parts in transition amplitude of meson decay
\cite{tH74}. The expression (\ref{2.12}) for $f_{\rho\pi\pi}$ is
just at the leading order of $1/N_c$ expansion, therefore, it must
be real. Thus $p^2<4m^2$ must be ensured in eq. (\ref{2.12}),
which means $m>m_\rho/2$ at $\rho$ energy scale. We should point
out that this conclusion is obtained from non-perturbative
expression (\ref{2.12}) which is characteristic of resummation
study.

The difference between our result of $m>385$MeV and the one of
$m=300$MeV in ref. \cite{Li95} is understandable. The
parametrization $m=300$MeV in ref. \cite{Li95} is consistent with
its $O(p^4)$ calculation, and the phenomenology studies are good
there. Now that we have resummed all terms in $p$-expansion, using
convergence or unitarity analysis, we have that $m>385$MeV. In
fact, such high constituent quark mass is also reasonable from
another aspect. When fitting the hadron spectra, we have to adopt
high constituent quark mass, or we fail to account for the
observed masses of scalar nonet \cite{Ripka}. It is argued in ref.
\cite{Ripka} that high constituent quark mass might be a
consequence of lack of confinement of the model and could be
avoided if we knew how to add confining forces to the model. This
is suggested by the results of constituent quark models which use
confining interactions \cite{confine}.

To conclude, we improve perturbative calculation in Li's model by
resummation study. We first illustrate proper vertex expansion
method in chiral quark model, and then use it to perform
resummation study on decay $\rho\rightarrow\pi\pi$. The complete
and compact expression of $f_{\rho\pi\pi}(p^2)$ (up to
$O(p^\infty)$) has been derived, from which the explicit
expression of effective action $S_{\rho\pi\pi}$ up to $O(p^8)$ is
easily obtained. We have seen that this method of resummation
study is a powerful method to catch informations from all orders
of $p$-expansion, from which we obtain non-perturbative conclusion
that $m>m_\rho/2$.

\begin{center}
{\bf ACKNOWLEDGMENTS}
\end{center}
This work is partially supported by NSF of China 90103002.

\end{document}